\documentclass[twocolumn]{aastex63}

\graphicspath{{./}{figures/}}

\newcommand{\gaia}{\textsl{Gaia}}

\shorttitle{Planet Occurrence Rate of Extragalactic Stars}
\shortauthors{Yoshida et al.}
\usepackage{physunits}
\usepackage{graphicx}
\usepackage{lineno}
\begin{document}

\title{Constraining the Planet Occurrence Rate around Halo Stars of Potentially Extragalactic Origin}

\author[0000-0003-4015-9975]{Stephanie Yoshida}
\affiliation{Department of Astronomy, Harvard University, 60 Garden Street, Cambridge, MA 02138, USA}

\author[0000-0003-4976-9980]{Samuel Grunblatt}
\altaffiliation{Kalbfleisch Fellow}
\affiliation{American Museum of Natural History, 200 Central Park West, Manhattan, NY 10024, USA}
\affiliation{Center for Computational Astrophysics, Flatiron Institute, 162 5th Avenue, Manhattan, NY 10010, USA}

\author[0000-0003-0872-7098]{Adrian~M.~Price-Whelan}
\affiliation{Center for Computational Astrophysics, Flatiron Institute, 162 5th Avenue, Manhattan, NY 10010, USA}

\begin{abstract}
The search for planets orbiting other stars has recently expanded to include stars from galaxies outside the Milky Way. With the TESS and \gaia\ surveys, photometric and kinematic information can be combined to identify transiting planet candidates of extragalactic origin.  Here, 1,080 low-luminosity red giant branch stars observed by \gaia\ and TESS with kinematics suggesting a high likelihood of extragalactic origin were searched for planet transits. Transit injection-recovery tests were performed to measure the sensitivity of the TESS data and completeness of the transit search. Injected signals of planets larger than Jupiter with orbital periods of 10 days or less were recovered in $\approx$44\% of cases. Although no planet transits were detected in this sample, we find an upper limit on planet occurrence of 0.52\% for hot Jupiters, consistent with previous studies of planet occurrence around similar host stars. As stars in the halo tend to be lower metallicity, and short period giant planet occurrence tends to be strongly correlated with stellar metallicity, we predict that relative to the Galactic disk population, a smaller fraction of halo stars will host planets detectable by transit surveys. Thus, applying the known planet occurrence trends to potential planet detection around halo stars, we predict $\gtrsim$7,000 stars must be searched with similar cadence and precision as the stars studied here before a detection of a planet of extragalactic origin is likely. This may be possible with future data releases from the TESS and \gaia\ missions.

\end{abstract}


\section{Introduction} \label{sec:intro}

To date, more than 4,000 extrasolar planets have been discovered, all of which have been relatively close to Earth \citep{akeson2013}. Most of these planets were found using the transit method in observations made by the Kepler  telescope \citep{borucki2011}. Kepler observed a small patch of sky toward the Galaxy's center for four years, revealing new features of the planet population at unprecedented precision, revealing the formation and evolution processes of planets of all sizes  \citep{fulton2017, weiss2018, berger2020}.

Although these discoveries have transformed our understanding of the Galaxy's planetary demographics, locating planets beyond the Local Group or even in other nearby galaxies remains controversial. The large difference between distances to stars in our Galaxy and stars in nearby galaxies have made learning about planets in other galaxies difficult requiring the development of unique detection techniques. Searches for planet candidates have been carried out in other galaxies via gravitational microlensing \citep{udalski2015}. However, due to the once-off chance alignments required for planet detection via gravitational microlensing, these candidates cannot be independently confirmed with other currently available techniques. Extragalactic planet candidates were also identified via X-ray eclipses \citep{distefano2020}. However, these candidates have yet to be verified by multiple independent studies. 

Additionally, the most well-characterized planet candidate of potentially extragalactic origin has been refuted. HIP 13044 is a star that passed its red-giant phase located in the Helmi Stream that merged with the Milky Way billions of years ago \citep{helmi2008}. In 2010, radial velocity observations made with Fibre-fed Extended Range Optical Spectrograph suggested the possibility of a planet orbiting this star \citep{setiawan2010}. However, reanalysis of the radial velocity data using a different approach did not show evidence for a planetary signal \citep{jones2014}.

In March 2018, NASA launched the Transiting Exoplanet Survey Satellite (TESS) with the goal of surveying the entire sky to discover hundreds or thousands of exoplanets \citep{barclay2018}. By utilizing the transit method, the telescope has already discovered over 180 exoplanets and more than 5000 planetary candidates. As the full-frame image data from TESS is publicly accessible, TESS data can be used to study a wide range of stellar samples across the sky. This is beneficial for observing luminous stars like giants, which are over-represented in a magnitude-limited dataset like that of TESS \citep{malmquist1922}.

In April 2018, the \gaia\ mission published the Data Release 2 (DR2), covering over 1 billion stars in the Galaxy. The mission of the \gaia\ space observatory is to measure precise stellar positions and space motions and understand the composition and evolution of the Milky Way \citep{gaia2018}.

A star population has been identified in the Galaxy’s inner halo with unique kinematic properties that appear to be distinct from the overall stellar halo population \citep{helmi2018}. The mostly retrograde rotational motion of these halo stars suggest the possibility of them being previously formed in a dwarf galaxy, now referred to as \gaia--Enceladus, which was presumed to have merged with the Milky Way 8 to 11 billion years ago \citep{myeong2019, grunblatt2021}. \gaia--Enceladus stars tend to show lower $\alpha$-element abundances than other halo stars \citep{nissen2010, hayes2018}. This implies that they have formed in a region with a low supernova rate relative to other halo stars, suggesting that they could not have formed in a galaxy with a high density of stars and dust like the Milky Way \citep{helmi2018}. However, since the elemental abundance information is not available for all of these targets, selections for this study were made based on kinematics and photometry only.

Here, we use \gaia\ DR2 data to isolate stars with kinematics consistent with an extragalactic origin, and search the corresponding TESS FFI data for planet transits. Additional cuts were made on stellar radius, determined in the TESS Input Catalog \citep{stassun2018}, to make a direct comparison to a previous study determining the planet occurrence around low-luminosity red giant branch stars with Kepler \citep{grunblatt2019}. Although no planet transits were found in our TESS sample, we performed an injection-recovery test to compute the completeness, and measure an upper limit on planet occurrence. This calculation allowed a comparison to planet occurrence around other star samples \citep{howard2012,grunblatt2019}.

\section{Target Selection} \label{sec:style}

The first target selection cuts were made in color and absolute magnitude using the \gaia\ DR2 data from sectors 1 to 26 with well-measured parallaxes and radial velocities to identify red giant stars \citep{grunblatt2019}. As giant stars are the most luminous, and most stellar halo stars lie at relatively large distances from the Sun, we exclude all stars with absolute \gaia\ $M_G$ $> 4.1$ and color \gaia\ $\textrm{B}_p-\textrm{R}_p < 0.97$ and $>3.0$, where absolute magnitudes have been determined by inverting \gaia\ parallaxes to compute distances. We did not use spectroscopic information in our target selection as spectroscopic information was only available for a subset of the stars in our initial target sample.

\begin{figure*}
\center
\includegraphics[width=0.48\textwidth]{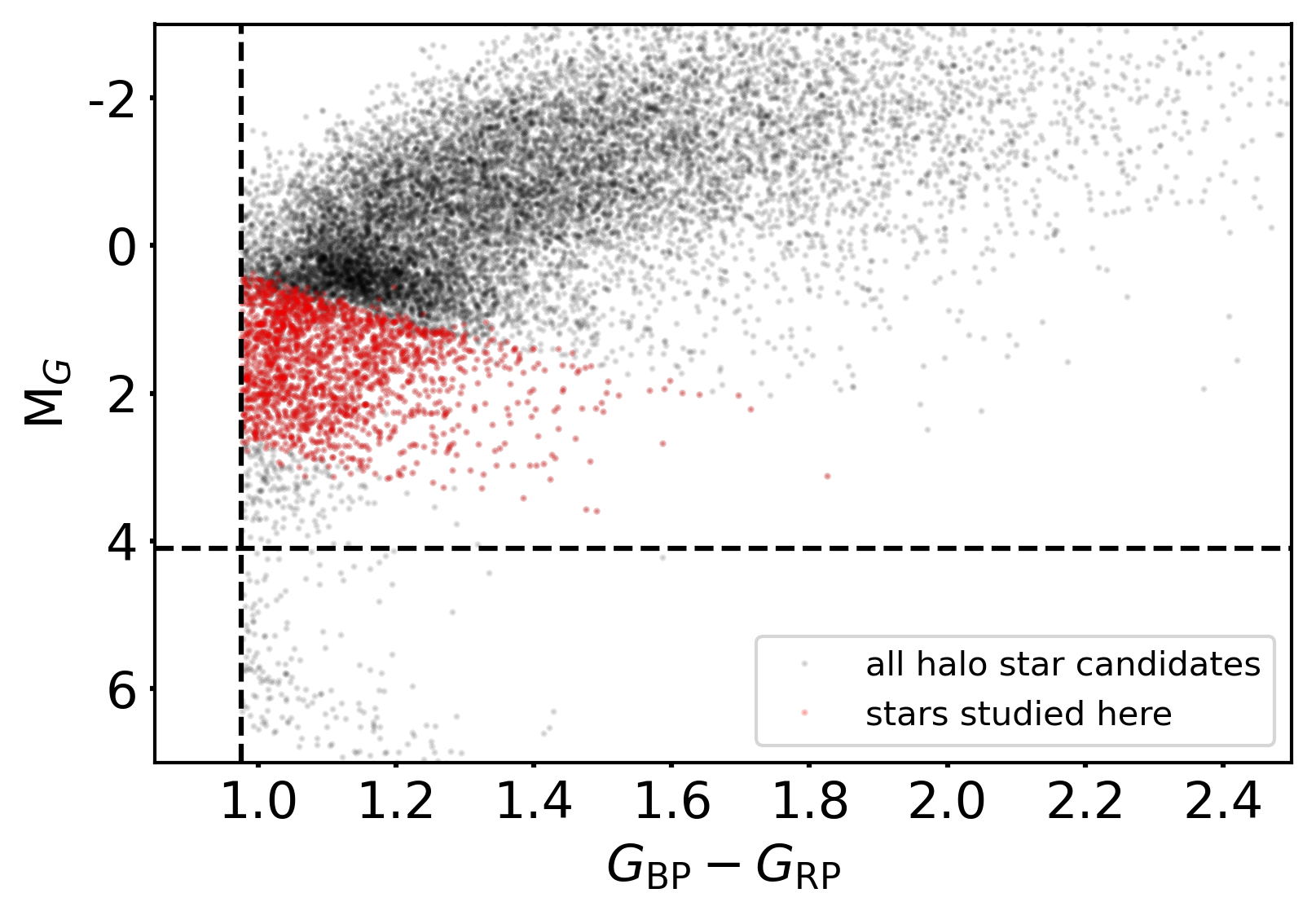}
\includegraphics[width=0.48\textwidth]{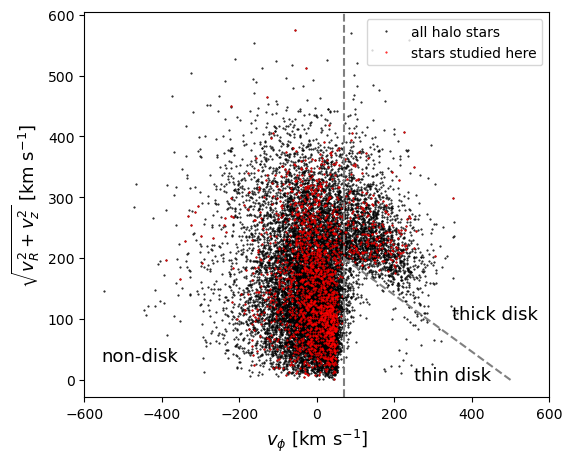}
\caption{\label{fig:fig1}Left: A color-magnitude diagram showing the absolute \gaia\ $m_G$ magnitudes and B$_p$-R$_p$ colors of stars in our sample. Our color and magnitude cuts correspond to the dashed lines. Further cuts were made based on stellar radius to focus on stars where transits are detectable, indicated by the red points. Right: A Toomre diagram of red giant stars in the Milky Way's halo and thick disk as a function of rotational and vertical velocity. The selection was made following the cuts in \citet{grunblatt2021}. Stars included in our injection-recovery test are shown as red points.}

\end{figure*}

A second reduction in target sample was made based on the stars' space motion. Because TESS is only sensitive to stars with relatively bright apparent magnitudes (m$_T$ $<$ 15), most of the giant stars in the photometric selection are likely to be kinematically associated with the Galactic disk. The stars selected for this study were low-luminosity red giants, all targets were at a TESS magnitude of 13 or less. To remove stars that rotate with the disk, velocities are computed in a Galactocentric reference frame, and then cuts in Galactic rotational and vertical velocity are used to remove Galactic disk stars following the procedure of \citet{grunblatt2021}. In particular, we convert heliocentric \gaia\ astrometry and radial velocity measurements \citep{lindegren2019,gaia2019} into Galactocentric cylindrical velocity components, $(v_R, v_\phi, v_z)$, using the \texttt{astropy} coordinate transformation framework \citep{astropy2013, astropy2018}. We use the default \texttt{v4.0} Galactocentric frame parameters,\footnote{This is a right-handed system with the Sun along the $-x$ axis and the solar velocity in the $+y$ direction.} which assume a solar Galactocentric distance of $8.122~{\rm kpc}$ \citep{gravity2018}, total solar velocity of $(-12.9, 245.6, 7.78)~\kmps$ \citep{drimmel2018, gravity2018, reid2004}, and height above the midplane of $20.7~\pc$ \citep{bennett2019}. We then remove stars that are associated with the kinematic thin and thick disk populations by only keeping stars with $v_\phi < 50~\kmps$ or $\sqrt{v_R^2 + v_z^2} > 200~\kmps$ (Figure \ref{fig:fig1}, right). For these stars, we then compute their orbital angular momenta ($z$-components, $L_z$) and energies ($E$, assuming the three-component Milky Way model implemented in \texttt{gala} \citep{gala}). To remove kinematic thick disk stars that pass our initial velocity selection, we additionally require $L_z > -600 \kmps~{\rm kpc}$ for our final sample of stellar halo giant stars. The distribution of these TESS--\gaia\ halo giants that remained after the reduction are a reasonable sampling of the local stellar halo, with a mean rotation near zero and a wide dispersion in vertical velocity.

Figure \ref{fig:fig1} shows the rotational and vertical Galactocentric velocities of stars that passed our color-magnitude and kinematic vetting as black and red points. In general, these targets have a low or negative rotational velocity unlike most stars that reside in the galactic disk. 


\subsection{TESS Pipeline} \label{subsec:tables}

After selecting our initial sample of targets based on \gaia\ measurements, we produced light curves using the Full Frame Image data from the NASA TESS Mission \citep{ricker2014}.

The 30-minute cadence full frame image data was converted into light curves for each target in our sample using the \texttt{giants} pipeline \citep{saunders2021}. This pipeline has been specifically designed to detect planets around giant stars by preserving longer-duration transits and transits near data gaps that have been diluted by other pipelines \citep{grunblatt2022}.

Additionally, we used the stellar radii determined by the TESS Input Catalog version 8 \citep{stassun2018}. We then used these stellar radii to make additional cuts to our sample, restricting our stellar radii to 3 to 8 R$_\odot$ to directly compare to planet occurrence rates of similar giant stars formed in our Galaxy \citep{grunblatt2019}. The 1,080 stars that passed these cuts are highlighted by the red points in Figure \ref{fig:fig1}.

\subsection{Transit Injection-Recovery Test}

After the target selection was completed, the initial sample of light curves was searched visually for planet transits. No strong signals were clearly identified, and thus an injection-recovery test was carried out to evaluate the completeness of our survey and establish upper limits of planet occurrence of this population.

Previous studies of planet occurrence determined through transit surveys of light curve data were able to justify their determined occurrence rate by evaluating their sensitivity to transit detection and completeness of their survey through the injection of synthetic transits and subsequent recovery of those signals using their detection algorithm \citep{christiansen2013,bryson2020}. In order to determine the sensitivity of our search to detect transiting planet candidates, we perform a more simplistic but similar injection-recovery test of the light curves tested here. Following selection of our fully detrended light curves using the size and color cuts described above, we injected the signal of model transits with a randomly assigned period and depth chosen through the process described in the following paragraph. We then attempt to recover this pure transit signal injected into the previously detrended light curves, and use our results from this search to determine our survey completeness.

\begin{figure*}[ht]
\centering
\includegraphics[width=0.48\textwidth]{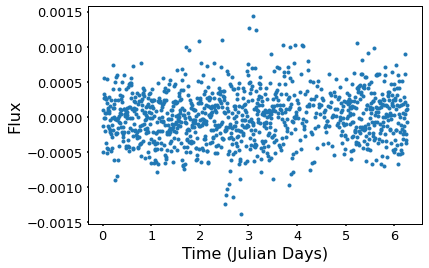}
\includegraphics[width=0.48\textwidth]{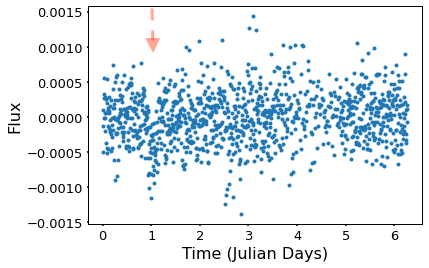}
\caption{Left: TIC455692967's light curve prior to the injection-recovery test. The target's flux is consistent throughout observation showing no indication of planetary transits. Right: TIC455692967's light curve following the injection of a 0.62 $R_J$ transit signal with a period of 6.26 days. The injected planet transit can be seen at a phase of 1 day in this figure indicated by the red arrow.}
\label{fig:fig2}
\end{figure*}

The simulated transits for this model were made with the Python package, \texttt{exoplanet} \citep{foreman2018}. The transit periods were drawn in a log-uniform distribution between three and fifty days and the planet to star radius ratios were drawn from a distribution between 0.001 and 0.045. The radius ratio and its respective range was used to mimic the injection-recovery process used in \citet{grunblatt2019}. The planet radii was determined from the measured transit depth and stellar radius from the TIC using equation 1:

\begin{equation}
\delta = (\frac{R_{\mathrm{p}}}{R_\star})^2
\end{equation}

where $R_p$ represents the planet radius, $R_*$ represents stellar radius, and $\delta$ represents transit depth. Figure \ref{fig:fig2} is a light curve undergoing the injection-recovery process. On the left is TIC455692967's light curve as a phase-folded plot prior to injection while on the right is the light curve following the injection of a 0.62 R$_\mathrm{J}$ planet with a 6.26 day period, representing a low signal-to-noise ratio injected signal that was successfully recovered.

Following the injection of the transit model into the target’s light curve flux, a Box Least Squares (BLS) search was run on each target \citep{kovacs2002}. The BLS search was used to look for transits with a period between three and fifty days within each light curve. Building on the logic of \citet{grunblatt2019}, we chose to consider a transit as recovered if the recovered period and depth are within 10\% of the injected period and 30\% of the injected depth, as visual vetting of a subset of recovered transits suggests that this period precision cut properly recovers $\gtrsim$85\% of all injected signals. To avoid the strongest systematic signals related to TESS' 13.7-day orbit, we rejected all signals with periods between 12.5 and 17 days from our injection-recovery test.


\section{Results}
\subsection{Survey Completeness}
After running 10,800 injection-recovery trials on our sample, the survey completeness and the sensitivity of the light curves to planet detection were determined.

\begin{figure*}[ht]
\centering
\includegraphics[width=0.24\textwidth]{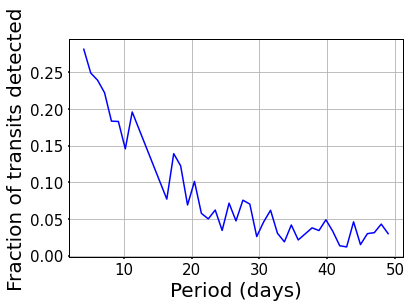}
\includegraphics[width=0.24\textwidth]{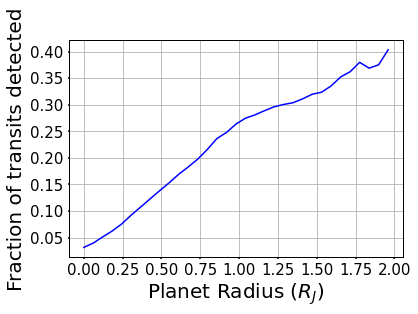}
\includegraphics[width=0.24\textwidth]{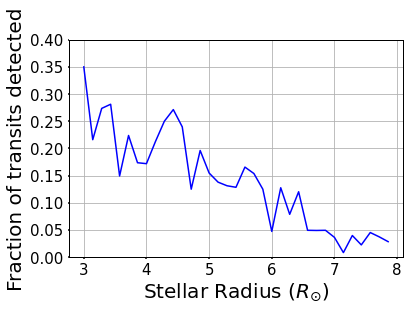}
\includegraphics[width=0.24\textwidth]{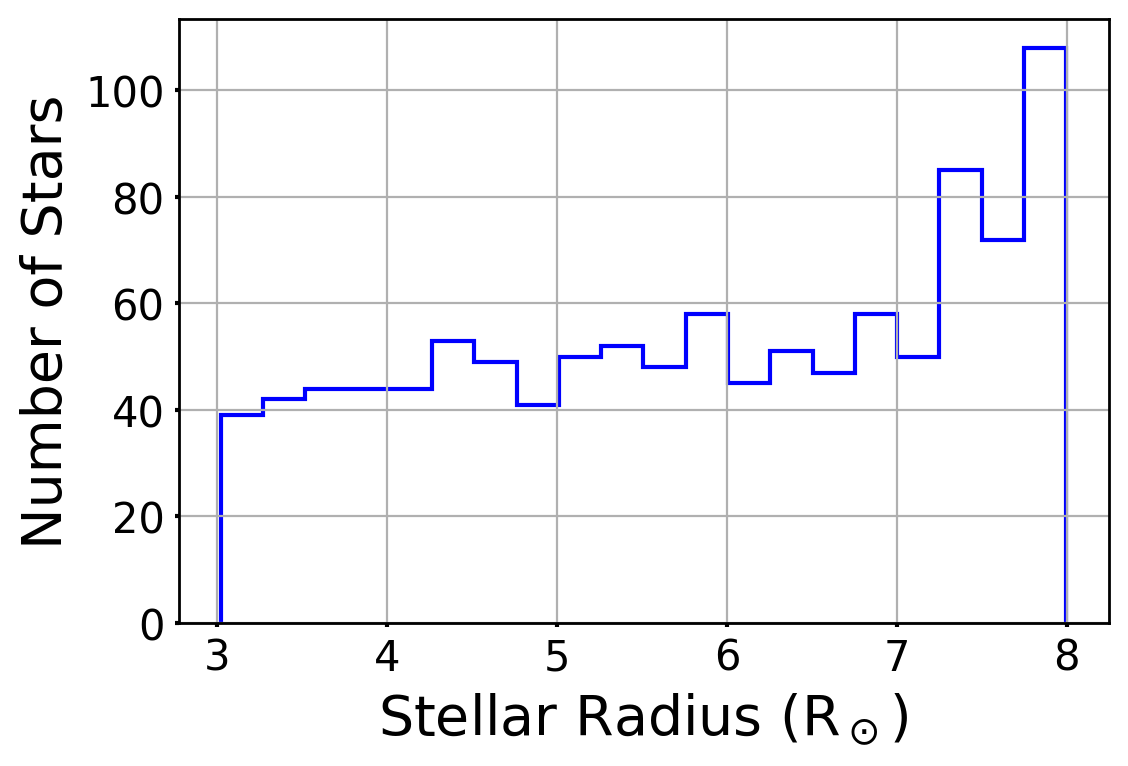}
\caption{Far Left: The recovery rate of the transits as a function of orbital period. Sensitivity to planet detection decreases as period increases. Left: The recovery rate of the transits as a function of planet radii. Sensitivity to planet detection increases as the planet radius increases. Right: The recovery rate of the transits as a function of stellar radius. The sensitivity has a negative correlation across all stellar radii sampled. Far right: The number of stars per stellar radius--there is an excess of large stars in the sample.}
\label{fig:fig3}
\end{figure*}

Figure~\ref{fig:fig3} displays the trend in the fraction of transits detected as the orbital period increases. We find that our BLS search recovered $\approx$50\% of injected transits at orbits of 5 days, and $\approx$10\% of injected transits at orbits of 30 days. Figure~\ref{fig:fig3}'s center-left panel shows the fraction of transits detected as a function of planet radius. We find that our BLS search recovered $\approx$30\% of injected transits of 1--2 R$_\mathrm{J}$ planets, and $\approx$15\% of $\leq$0.5 R$_\mathrm{J}$ planets. Figure \ref{fig:fig3}'s center-right panel shows the fraction of planets detected as a function of the stellar radius. We find that $\approx$30\% of injected signals are detected around 3 R$_\odot$ stars, while $<$5\% of signals are detected around 7 R$_\odot$ stars. Figure \ref{fig:fig3}'s rightmost panel shows the number of stars in our sample as a function of stellar radius. $>$50\% of the stars in our sample had radii of $>$ 5.8 R$_\odot$. We predict that completeness would be higher if a higher fraction of smaller radius stars were tested in this study.

\begin{figure*}[ht]
\centering
\includegraphics[width=0.48\textwidth]{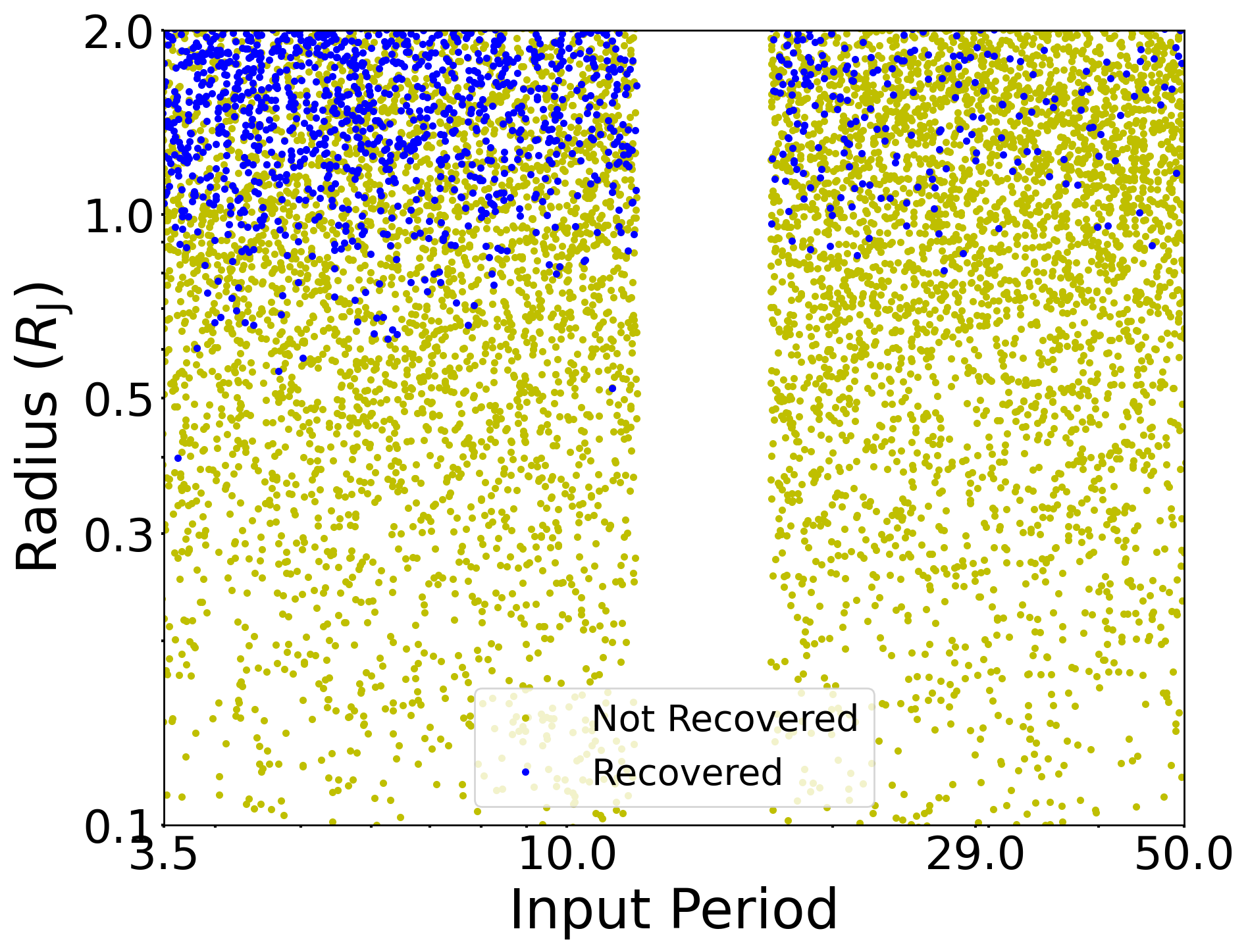}
\includegraphics[width=0.48\textwidth]{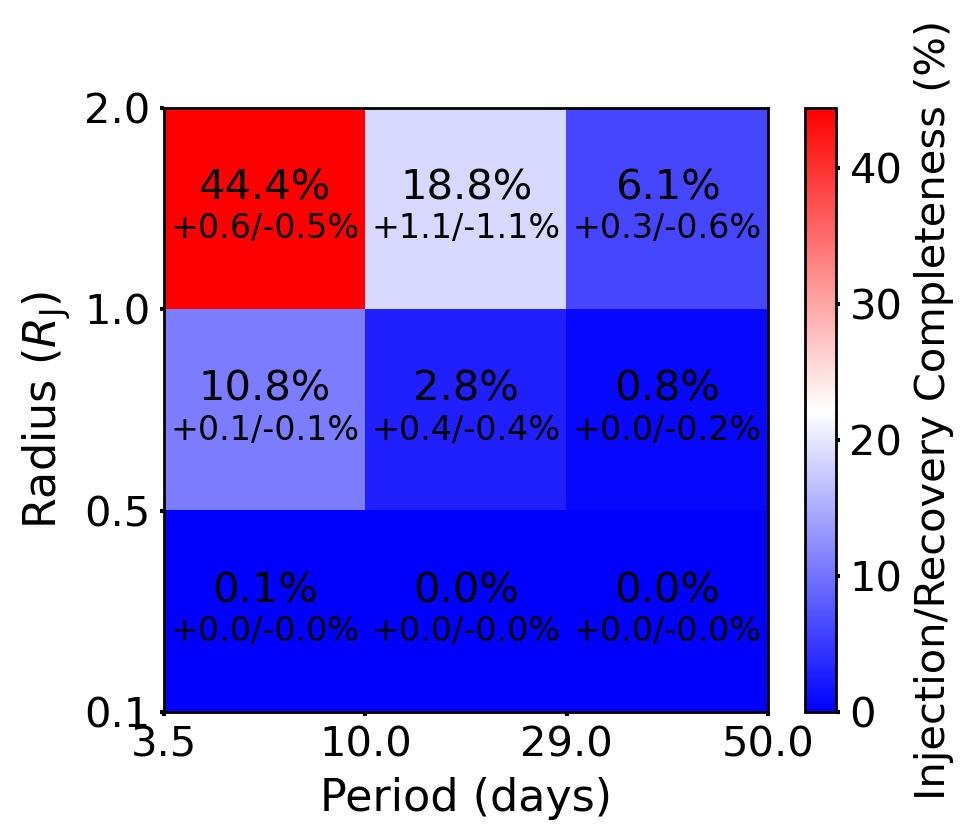}
\caption{Left: The transit signals injected into the observed halo stars as a period by radius plot. The blue indicates the targets recovered and the orange indicates the targets that were not successfully recovered. The gap down the middle is from the masked period range of 12.5 to 17 days. Right: The recovery rate of the transits as a function of planet period and radius. Completeness is highest for largest and shortest period planets.}
\label{fig:fig4}
\end{figure*}

Figure \ref{fig:fig4} reveals the distribution of ``recovered'' and ``unrecovered'' transit signals. Out of the 10,800 injection-recovery trials, only 2,987 injected signals were recovered by the BLS search. As seen in Figure \ref{fig:fig4}, the majority of the recovered targets are concentrated in the top left corner of the plot, illustrating that targets with a larger planet radius and smaller orbital period are more likely to be detected. 

The completeness was evaluated further after being separated into nine different bins (Figure \ref{fig:fig4}) as:

\begin{equation}
c_\mathrm{bin} =  n_\mathrm{trans,det/bin}/n_\mathrm{trans/bin}
\end{equation}

where $n_\mathrm{trans,det/bin}$ is the number of transiting planets detected per bin, $n_\mathrm{trans/bin}$ is the number of simulated planets per bin, and $c_\mathrm{bin}$ is the completeness fraction per bin. These bins were chosen for direct comparison to previous transit-based planet occurrence studies \citep{howard2012, grunblatt2019}. To determine errors for the injection-recovery test, we adjusted our criteria to consider transits ``recovered'' if the BLS-measured best fit period agreed to within 15\% of the injected period to determine upper uncertainties, and within 5\% for lower uncertainties. In all cases, the visual evidence for any recovered injected signal is stronger than the signal of any potential planet candidates identified in the initial search. To test this, we also injected a transit signal with the same transit depth and period as our best planet candidate signal identified in our initial search. Comparable signals were not detected in $\gtrsim$85\% of injection-recovery test cases, implying that our visual inspection of the data is similarly complete as our subsequent transit injection-recovery test, and none of the signals identified in our visual search warrant designation as transiting planet candidates.

We find that completeness is only $>$ 20\% for the largest, shortest period planets (3.5 $<$ P $<$ 10 d, 1.0 $<$ R$p$ $<$ 2.0 R$_\mathrm{J}$). Injected transits are recovered for similarly sized planets on slightly longer orbits (10 $<$ P $<$ 29 d, 1.0 $<$ R$p$ $<$ 2.0 R$_\mathrm{J}$) in one fifth of cases. Any injected planet with a radius below 0.5 $R_{\mathrm{J}}$, regardless of orbital period, was only detected in one case, for a signal with a period $<$ 4 days. 

There are potentially multiple sources of error that could have affected the recovery rate of the injected transits. The TESS telescope has systematic noise due to detector and instrumental issues and scattered background light that could have affected the light curves' signal-to-noise ratio. The fluctuation in noise levels could have changed the telescope's ability to detect the planet's real signal--the BLS search would recognize the noise as the best fit period rather than the planetary signal. These systematic signals are strongest at or near the TESS data downlink 13.7 day period, motivating our choice to not inject any signals with periods between 12.5 and 17 days (Figure \ref{fig:fig4}). In addition, the principal component analysis approach of the \texttt{giants} pipeline was not always successful at removing all systematic noise, and in some cases could not remove the data's largest noise sources \citep{saunders2019}. The largest sources of noises could be instrumental noise, statistical fluctuations, poor detrending, and/or stellar variability \citep{thompson2018}. Given that the systems observed by TESS in this project were ultimately those with large apparent magnitudes, their signals may have been polluted by systematic signals from either astrophysical or instrumental sources. This could cause multiple targets to lie within the same or adjacent pixels making source confusion an issue. The absence of planets between 12.5 and 17 days may have also biased the recovery rates for the bins with periods between 10 and 29 days as planets used for a portion of the range could not be injected or recovered. The analysis presented here follows the procedure described in \citet{grunblatt2019}, with validation of the injection-recovery process carried out with a similar visual inspection (both the smallest-depth and longest-period injected signals were visually verified to ensure a transit would have been identified through visual inspection of the light curve).




\section{Discussion}



\subsection{Upper Limits on Planet Occurrence}
In table 1, calculations for upper limits of planet occurrence were made for each bin in our completeness figure. As there were no transits found by our BLS search in this sample, the planet occurrence was computed by assuming that there is less than one planet per bin. The planet occurrence per bin, f$_\mathrm{bin}$, is determined to be

\begin{equation}
    f_\mathrm{bin} = \frac{n_\mathrm{pl,aug,bin}}{n_\star \times c_\mathrm{bin}}
\end{equation}

where $n_\mathrm{pl,aug,bin}$ represents the augmented number of planets found in a bin, $n_\star$ represents the number of stars in the sample, and $c_\mathrm{bin}$ represents each bin's completeness, given by the injection-recovery test. The augmented number of planets is determined by multiplying the maximum number of planets in a bin (1) by the ratio of the median bin orbit size to the median stellar radius:

\begin{equation}
    n_\mathrm{pl,aug,bin} = 1 \times \frac{a_\mathrm{bin,med}}{R_{\star,\mathrm{med}}}
\end{equation}

where the median orbital distance for a planet in a given bin is represented as $a_\mathrm{bin,med}$ and the median size of a star in this sample and the median size of a star in a bin is given as $R_{\star,\mathrm{med}}$: for bins from 3.5 to 10 days, $R_{\star,\mathrm{med}}$ is 2.5; 10 to 29, $R_{\star,\mathrm{med}}$ is 6; for 29 to 50, $R_{\star,\mathrm{med}}$ is 12. Our determination of an occurrence rate of $<$ 0.52 \% for hot Jupiters (the largest-radius, smallest-period bin) orbiting low-luminosity red giant halo stars using TESS agrees with the \citet{grunblatt2019} estimate of 0.51\% $\pm$ 0.29\% of low-luminosity red giant stars hosting hot Jupiters. 



\begin{deluxetable*}{ccCcrlc}
\tablecaption{Upper Limits on Planet Occurrence \label{tab:mathmode}}
\tablecolumns{6}
\tablenum{1}
\tablewidth{0pt}
\tablehead{
\colhead{Period vs. Planet Radius} &
\colhead{3.5 to 10 days} &
\colhead{10 to 29 days} &
\colhead{29 to 50 days}
}
\startdata
1.0 to 2.0 $R_{\mathrm{J}}$ & 0.52\% & 2.96\% & 18.2\% \\ 
& {\small +0.01/-0.01\%} & {\small +0.18/-0.16\%} & {\small +0.85/-1.99\%} \\ 
0.5 to 1.0 $R_{\mathrm{J}}$ & 2.14\% & 27.7\% & {\it undefined}\* \\
& {\small +0.02/-0.02\%} & {\small +10.4/-4.63\%} &  \\
0.1 to 0.5 $R_{\mathrm{J}}$ & {\it undefined}\* & {\it undefined}\* & {\it undefined}\* \\
\enddata
\caption{Note: No injected planets were recovered with radii $<$0.5 R$_\mathrm{J}$ so completeness is 0\% and thus we present the planet occurrence rate as undefined.}
\end{deluxetable*}



\subsection{Effects of Stellar Metallicity}
Stellar metallicity is known to correlate strongly with the occurrence of hot Jupiters, which is proportional to $10^{[{\rm Fe}/{\rm H}] \times 1.2}$ \citep{fischer2005,johnson2010,ghezzi2018}. Furthermore, halo stars are known to typically have metallicities of $[{\rm Fe}/{\rm H}] = -0.8$ or lower \citep{mackereth2019}. This implies a factor of $\sim10$ decrease in planet occurrence for halo stars relative to stars in the rest of our Galaxy. Therefore, at least 10 times as many stars must be observed with the quality obtained by this survey to distinguish between metallicity and galactic origin effects. Other injection-recovery tests of low-metallicity TESS stars beyond the Galactic thin disk suggest that the hot Jupiter occurrence rate around these stars is $<$0.18\%, significantly lower than the planet occurrence rate around solar-metallicity stars \citep{boley2021}.



Furthermore, there are stars that originated in the Milky Way but were mixed into this halo population by dynamical interactions with infalling galaxies such as Gaia--Enceladus. However, these {\it in situ} stars likely represent a small fraction of stars in our sample, as these stars tend to have higher rotational and lower vertical velocities compared to the majority of stellar halo stars. The combination of spectroscopic and kinematic surveys of halo K-giant stars suggest that stars formed {\it in situ} may represent $\sim 3\%$ of the stars which passed the rotational and vertical velocity cuts made by this study \citep{xue2014, belokurov2020}. Furthermore, the presence of a larger {\it in situ} halo population would reduce the slope of accreted metallicity-stellar mass relationship for Milky Way-like mass galaxies, in conflict with the observed relationship between mass and metallicity of stellar halos in Milky Way-mass spiral galaxies \citep{harmsen2017, dsouza2018}.

\section{Conclusion and Future Investigation}

By cross-matching the datasets from \gaia\ DR2 and TESS full frame images from sectors 1 to 26, we selected and searched a red giant star sample with negligible or retrograde Galactic motion for planet transits. We analyzed light curves produced from TESS full frame image data of low-luminosity giant stars with kinematic properties similar to the Galactic thick disk and halo. After running a BLS search on this light curve sample, no planet candidates were identified, most likely due to the limited sample size (1,080 stars). Using these light curves, a transit injection-recovery test was done to determine the completeness of this planet transit survey. The injection-recovery test results indicate that planets with orbital period  $<$10 days and $R_p > R_J$  are the only planets that were recovered in $>$40\% of light curves. Using these completeness estimates, an upper limit of planet occurrence was estimated for this population, indicating $<0.5\%$ of the stars in this sample likely host hot Jupiters (3.5 to 10 day period, $>1$ $R_{\mathrm{J}}$). This is in agreement with previous estimates of planet occurrence around red giants and planet occurrence estimates around metal-poor stars \citep{mortier2012,grunblatt2019,boley2021}.  Considering the  planet occurrence found for similar stars in our own Galaxy, effects of metallicity and kinematic mixing in the halo, and assuming planet occurrence is similar in other galaxies, we expect a number of targets likely needed to be searched $n_\mathrm{halo} \gtrsim \frac{1}{f_\mathrm{RG} (R_*/a)k_\mathrm{KM} k_\mathrm{[Fe/H]}}$ where $f_\mathrm{RG}$ is the occurrence rate of hot Jupiters orbiting red giants \citep[$0.51\%$]{grunblatt2019}, $(R_*/a)$ the average transit probability for these hot Jupiters (we assume 0.3), k$_\mathrm{KM}$ is the effect of kinematic mixing ($\approx$0.9), and k$_\mathrm{[Fe/H]}$ is the factor of metallicity on planet occurrence ($\approx0.1$). We find that $n_\mathrm{halo} \gtrsim 7,000$ targets selected via our cuts observed with similar precision may reveal a transiting planet of extragalactic origin, in agreement with our findings of 0 planet candidates among 1,080 targets. \emph{Gaia} DR3, released earlier this year \citep{gaiadr3rvs}, has provided radial veloicites for fainter stars that were excluded from this survey. This, along with Gaia DR3 parallaxes, could allow identification of $\sim$5 times as many halo stars.


\acknowledgements{We acknowledge Daniel Huber for helpful discussions. We acknowledge the use of public TESS data from pipelines at the TESS Science Office and at the TESS Science Processing Operations Center. Resources supporting this work were provided by the NASA High-End Computing (HEC) Program through the NASA Advanced Supercomputing (NAS) Division at Ames Research Center for the production of the SPOC data products. S.G. acknowledges support by the National Aeronautics and Space Administration under Grants 80NSSC19K0593 and 80NSSC21K0781 issued through the TESS Guest Investigator Program. Funding for the TESS mission is provided by NASA's Science Mission Directorate.}

\end{document}